\documentclass[reprint, superscriptaddress, secnumarabic, amssymb, nobibnotes, aps, prl]{revtex4-1}

\usepackage{graphicx}
\usepackage{epstopdf}
\usepackage[T1]{fontenc}
\usepackage[latin9]{inputenc}
\usepackage{amsbsy}
\usepackage{gensymb}
\usepackage[T1]{fontenc}
\usepackage[latin9]{inputenc}
\usepackage{amsmath}
\usepackage{amssymb}
\usepackage{bbm}
\usepackage{physics}
\usepackage{braket}
\usepackage{xcolor}
\allowdisplaybreaks
\usepackage{graphicx}
\usepackage[colorlinks=true]{hyperref}  
\hypersetup{
    bookmarks=true,         
    unicode=false,          
    pdftoolbar=true,        
    pdfmenubar=true,        
    pdffitwindow=false,     
    pdfstartview={FitH},    
    pdftitle={unconventional superconducting properties in noncentrosymmetric superconductor Re5.5Ta},    
    pdfauthor={Arushi, D. Singh, P. K. Biswas, A. D. Hillier, R. P. Singh},     
    pdfsubject={},   
    pdfcreator={},   
    pdfproducer={}, 
    pdfkeywords={} {} {}, 
    pdfnewwindow=true,      
    colorlinks=true,       
    linkcolor=blue, 
    citecolor=blue,        
    filecolor=magenta,      
    urlcolor=blue           
} 
\usepackage[normalem]{ulem}

\newcommand{\equref}[1]{Eq.~(\ref{#1})}

\newcommand{\figref}[1]{Fig.~\ref{#1}}

\begin{document}
\title{\textrm{Unconventional superconducting properties of noncentrosymmetric Re$_{5.5}$Ta}}
\author{Arushi}
\affiliation{Department of Physics, Indian Institute of Science Education and Research Bhopal, Bhopal, 462066, India}
\author{D.~Singh}
\affiliation{ISIS Facility, STFC Rutherford Appleton Laboratory, Didcot OX11 0QX, United Kingdom}
\author{P.~K.~Biswas}
\affiliation{ISIS Facility, STFC Rutherford Appleton Laboratory, Didcot OX11 0QX, United Kingdom}
\author{A.~D.~Hillier}
\affiliation{ISIS Facility, STFC Rutherford Appleton Laboratory, Didcot OX11 0QX, United Kingdom}
\author{R.~P.~Singh}
\email[]{rpsingh@iiserb.ac.in}
\affiliation{Department of Physics, Indian Institute of Science Education and Research Bhopal, Bhopal, 462066, India}
\date{\today}

\date{\today}
\begin{abstract}
\begin{flushleft}
\end{flushleft}
Rhenium based noncentrosymmetric superconductors crystallizing in $\alpha$-Mn structure have become potential candidates to exhibit an unconventional superconducting ground-state. Here we report a detailed investigation on the superconducting and normal state properties of Re$_{5.5}$Ta, that also has the $\alpha$-Mn structure. Magnetization, specific heat, and transport measurements confirm the bulk superconducting transition \textit{T}$_{C}$ at 8.0 K. Upper critical field value (H$_{C2}$(0)) calculated from magnetization, specific heat and AC transport measurements exceed the Pauli paramagnetic limit (14.7 T), indicating that the superconducting properties of Re$_{5.5}$Ta are probably unconventional in nature. However, low-temperature specific heat and transverse-field muon spin rotation measurements suggest a surprising nodeless isotropic superconducting gap, although with strong electron-phonon coupling.
\end{abstract}
\maketitle
\section{Introduction}

In the superconducting state, the symmetry of the order parameter plays a vital role. In conventional superconductors, gauge symmetry is broken, whereas additional symmetries, such as time-reversal symmetry may also be broken in unconventional superconductors. Non-centrosymmetric (NC) superconductors have recently emerged as an exciting class of unconventional superconductors, where the underlying crystal structure lacks inversion symmetry \cite{CPS,CRS_1,CRS_2,LP_PB,KCA,BP1,MIB}. The lack of inversion symmetry in these superconductors removes the spin degeneracy of the electronic states through an anti-symmetric spin-orbit coupling (ASOC). It leads to the splitting of the Fermi surface into two different helicity bands and has a strong influence on the possible Cooper pairing states. This, in turn, may induce the admixture of spin singlet and triplet components where the extent of mixing is determined by the magnitude of ASOC splitting \cite{Sing-Trip_1,Sing-Trip_2}. This possible mixed pairing may lead to superconductors with exotic properties which are not observed in conventional superconductors, such as: upper critical field exceeding or close to Pauli paramagnetic limit \cite{CPS,Pauli1,Pauli2}, nodes in the superconducting gap, \cite{LPB_nod,YC_nod,LC_2gap,LNC_2gap} and time-reversal symmetry breaking (TRSB) which is a rarely observed phenomenon \cite{Zr_TRS,Hf_TRS,R6T,R4.8T,Re-Nb,LI,LNC_TRS}. 

Recently, Re-based superconducting binary alloys [Re\textit{X} (\textit{X} = transition metal)] \cite{Re-Nb, Zr_TRS,Hf_TRS,R6T,R4.8T} crystallizing in NC $\alpha$-Mn structure have received huge attention due to the frequent occurrence of TRSB in this series of compounds. But the similar TRSB signal below T$_{C}$ in these materials, independent of transition metal \textit{X} raises a question about the role played by the strength of spin-orbit coupling. Recent observation of TRSB in pure Re, which crystallizes in a centrosymmetric structure with space group: P6$_{3}$/\textit{mmc} \cite{Re-Nb} indicates that the dominant d-bands of Re and its local electronic structure may be crucial for the TRSB in Re based superconductors. At the same time, the absence of TRSB in Re$_{3}$W \cite{R3W_1,R3W_2} and Re$_{3}$Ta \cite{R3T} despite its crystallization in a $\alpha$-Mn non-centrosymmetric structure raises further questions. It suggests the presence of a critical amount of Rhenium which gives rise to the exotic features in the superconducting state, rather than the inversion asymmetry. To further understand the role of Re in time-reversal symmetry breaking in elemental Re and Re-based compounds, it is clearly required to study new Re rich compounds that crystallize in $\alpha$-Mn crystal structure.

In this paper, we report the synthesis of a new compound Re$_{5.5}$Ta having a NC $\alpha$-Mn crystal structure. The magnetization, resistivity, and specific heat measurements confirm bulk type-II superconductivity with an onset temperature of 7.95 K. The upper critical field value determined by the previously mentioned measurements exceed the Pauli paramagnetic limit (14.7 T). This is a strong indicator for possible unconventional superconductivity. Muon spin rotation and relaxation ($\mu$SR) measurements provide the information on the superconducting gap symmetry and nature of the superconducting ground state.   
\ 
\section{Experimental Details}
A polycrystalline sample of Re$_{5.5}$Ta was prepared by arc melting the constituent high purity elements, Re and Ta on a water-cooled copper hearth under an argon gas atmosphere. The as-cast ingot was flipped and remelted several times to ensure phase homogeneity with negligible weight loss. A powder X-ray diffraction (XRD) pattern was collected at room temperature using a PANalytical diffractometer equipped with CuK$\alpha$ radiation ($\lambda$ = 1.5406 \AA). Magnetic, specific heat, and electrical resistivity measurements were done on a superconducting quantum interference device (MPMS 3, Quantum Design) and a 9T physical property measurement system (PPMS). Muon spin rotation/relaxation measurements were performed on the MuSR spectrometer \cite{Muon} at the ISIS pulsed muon facility, Rutherford Appleton Laboratory, United Kingdom. The measurements were performed in zero-field (ZF), transverse-field (TF), and longitudinal-field (LF) configurations.

\begin{figure} 
\includegraphics[width=1.0\columnwidth, origin=b]{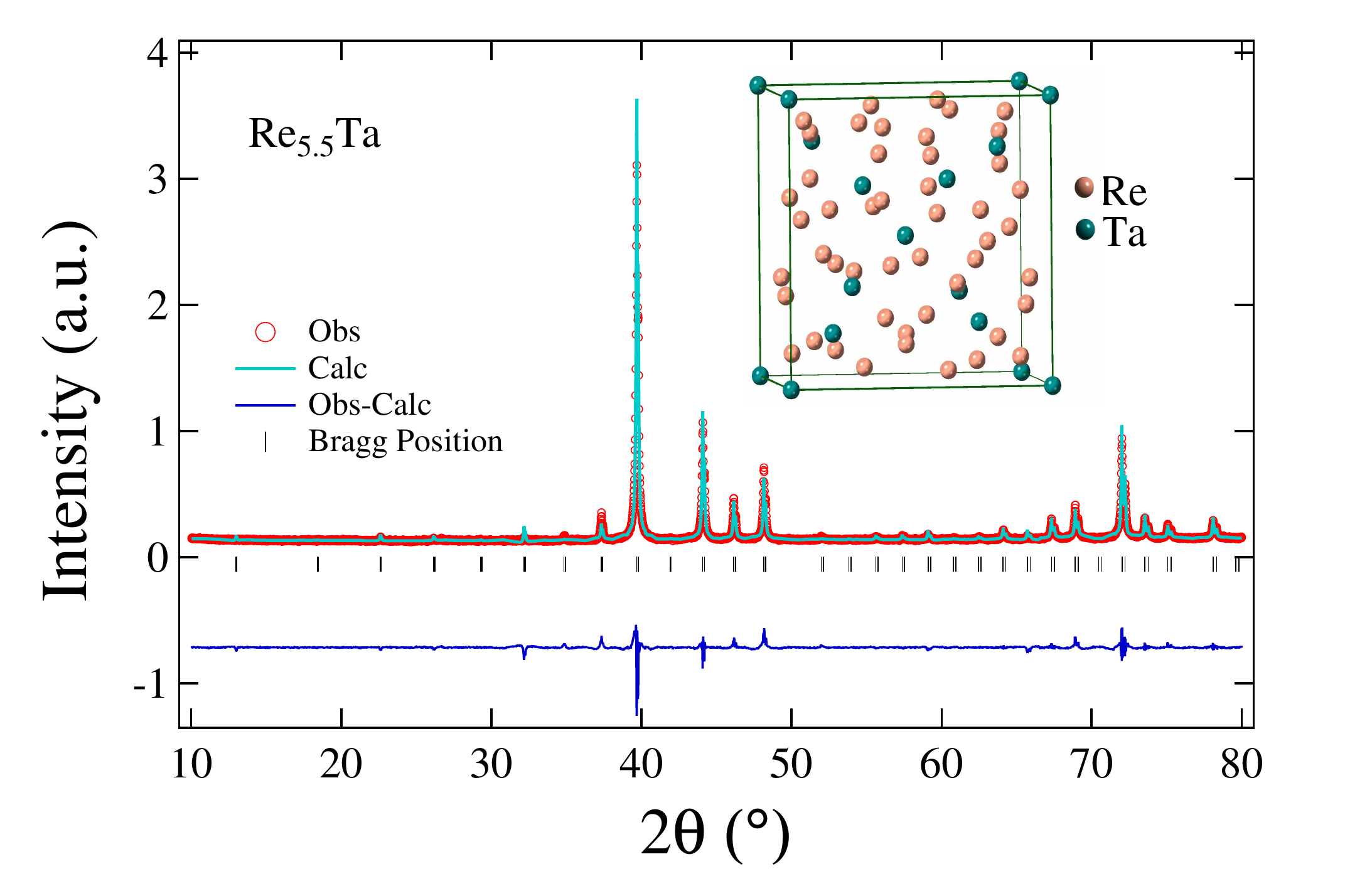}
\caption{\label{Fig1:XRD} Room temperature powder XRD pattern obtained for Re$_{5.5}$Ta sample is shown by open red circles. The solid green line represents the Rietveld refinement whereas the black bars show Bragg reflection peaks. The difference between the observed and calculated patterns is shown by a blue line. Inset: Crystal structure of Re$_{5.5}$Ta.}
\end{figure}

\section{Results and Discussion}
\subsection{Sample characterization}
The XRD pattern collected at room temperature was refined using the FullProf software. The refinement, shown in \figref{Fig1:XRD}, confirms the phase purity and the crystal structure as cubic $\alpha$-Mn with space group I$\bar{4}3$\textit{m}. The lattice constant is: a = 9.628 \text{\AA}. Other parameters obtained from the refinement, such as occupancy, atomic and Wyckoff positions, are listed in Table I. The crystal structure of Re$_{5.5}$Ta corresponding to one unit cell is given in the inset of \figref{Fig1:XRD} where Ta atoms are represented by solid green circles and Re atoms by solid orange circles.
\begin{table}[h!]
\caption{Structure parameters of Re$_{5.5}$Ta obtained from the Rietveld refinement of XRD}
\begin{tabular}{l r} \hline\hline
Structure& Cubic\\
Space group&        I$\bar{4}$3\textit{m}\\ [1ex]
Lattice parameters\\ [0.5ex]
a (\text{\AA})&  9.628(3)\\
V$_{Cell}$ (\text{\AA})& 892.90(3)
\end{tabular}
\\[1ex]

\begingroup
\setlength{\tabcolsep}{4pt}
\begin{tabular}[b]{c c c c c c}
Atom&  Wyckoff position& x& y& z& Occupancy\\[1ex]
\hline
Ta1& 8c& 0& 0& 0& 0.4757\\             
Ta2& 2a& 0.324& 0.324& 0.324& 0.0799\\                       
Re1& 24g& 0.359& 0.359& 0.042& 0.4446\\
Re2& 24g& 0.086& 0.086& 0.279& 0.4698\\
[1ex]
\hline
\end{tabular}
\par\medskip\footnotesize
\endgroup
\end{table}

\subsection{Superconducting and normal state properties}

\begin{figure} 
\includegraphics[width=1.0\columnwidth, origin=b]{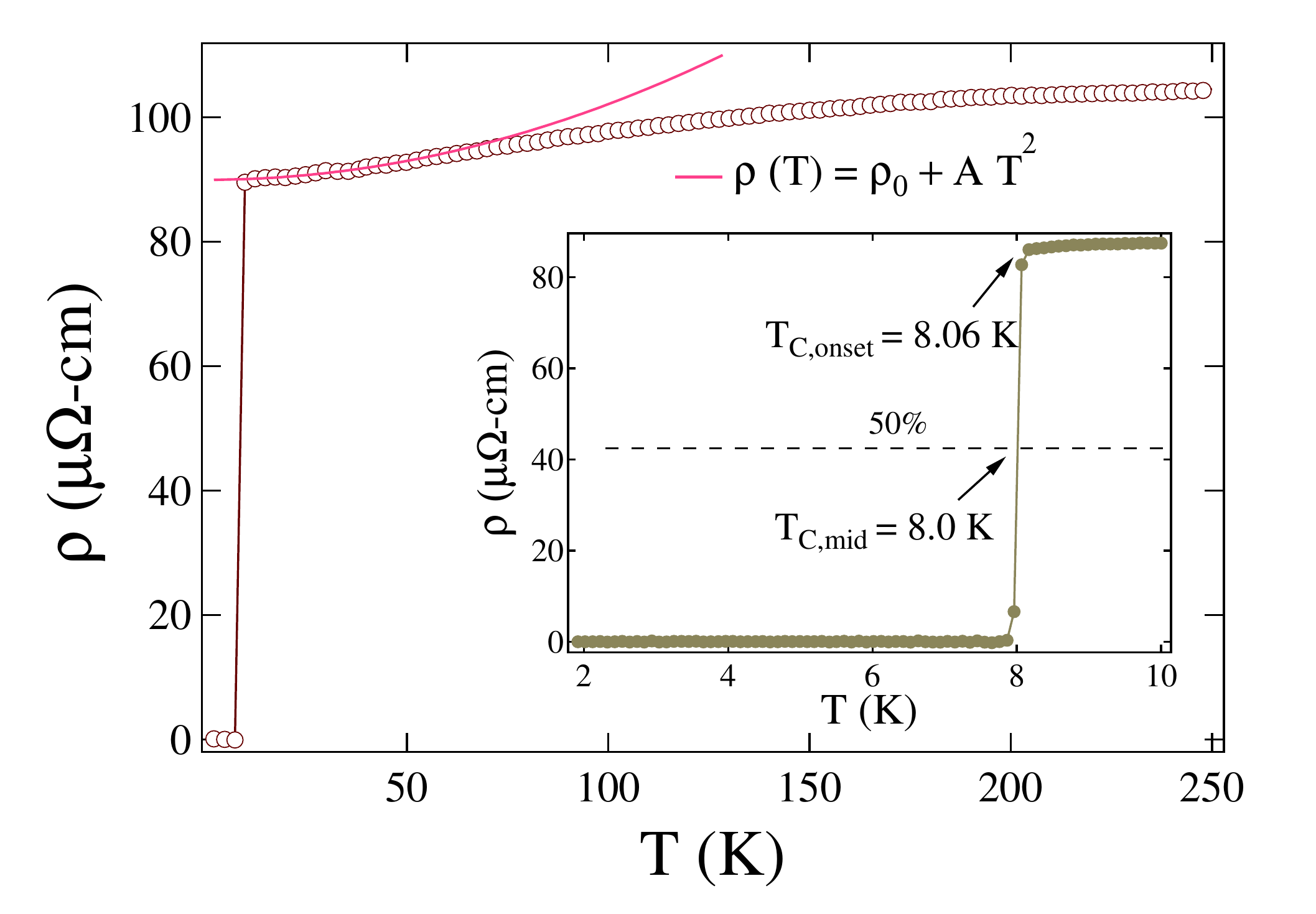}
\caption{\label{Fig2:Res} Temperature dependence of resistivity at zero field over the temperature range 1.8 K $\leq$ T $\leq$ 250 K, where the data above T$_{C}$ up to 50 K is well fitted by power law: $\rho(T) = \rho_{0} + AT^{2}$, shown by solid line. Enlarged view of $\rho$(T) data exhibiting superconducting transition at T$_{C,mid}$ = 8.0 K.}
\end{figure}

\subsubsection{Electrical resistivity}
Electrical resistivity versus temperature, $\rho$(T), measurement was performed in zero applied magnetic field in the temperature range 1.8 K $\leq$ T $\leq$ 250 K [see \figref{Fig2:Res}]. The residual resistivity ratio, found to be 1.2 for Re$_{5.5}$Ta is in good agreement to other Re based materials \cite{FerLiqd_3&Re6Zr_WTRS, R24T5_WTRS, Fer_Liqd_1&N0.18R0.82}.

$\rho(T)$ drops to zero at T$_{C,onset}$ = 8.06 K, resulting in T$_{C,mid}$ = 8.0 K and 90 - 10$\%$ transition width $\Delta$T < 0.2 K [see inset of \figref{Fig2:Res}]. The low-temperature resistivity data were fitted in the temperature range 10 K to 50 K using a power-law relation: $\rho(T) = \rho_{0} + AT^{n}$, where $\rho_{0}$ represents the residual resistivity due to impurities and lattice defects, and A gives information about the degree of electron-electron correlation in the material. The best fit to the data is obtained when n = 2, which suggests the Fermi-liquid type temperature dependence in the normal state resistivity \cite{FerLiqd_3&Re6Zr_WTRS, Fer_Liqd_1&N0.18R0.82, Fer_Liqd_2}, yielding $\rho_{0}$ = 89.95 $\pm$ 0.08 $\mu\Omega $-cm and A = 0.00121 $\pm$ 0.000064 $\mu\Omega $-cm K$^{-2}$. The values of A and $\rho_{0}$ suggests that Re$_{5.5}$Ta is a weakly correlated and disordered system, respectively.

The Kadowaki-Woods ratio is a measure of the magnitude of the electron-electron correlation and is given by the expression K$_{w} = A/\gamma_{n}^{2}$ where A is the coefficient of quadratic term in resistivity and $\gamma_{n}$ is given by Sommerfeld constant from specific heat data. By taking A = 0.00121 $\mu\Omega $-cm K$^{-2}$ and $\gamma_{n}$ = 25.33 mJmol$^{-1}$K$^{-2}$, the value obtained for K$_{w}$ = 0.18$\times$10$^{-5}$ $\mu\Omega $-cm mJ$^{-2}$mol$^{2}$ K$^{2}$ is significantly smaller than the value obtained in strongly correlated systems, such as heavy Fermion compounds (1.0$\times$10$^{-5}\mu\Omega$-cm mJ$^{-2}$ mol$^{2}$ $K^{2}$) \cite{KWR_1,KWR_2,KWR_3}, which suggests that Re$_{5.5}$Ta is a weakly correlated system.

\subsubsection{Magnetization}
\begin{figure} 
\includegraphics[width=1.0\columnwidth, origin=b]{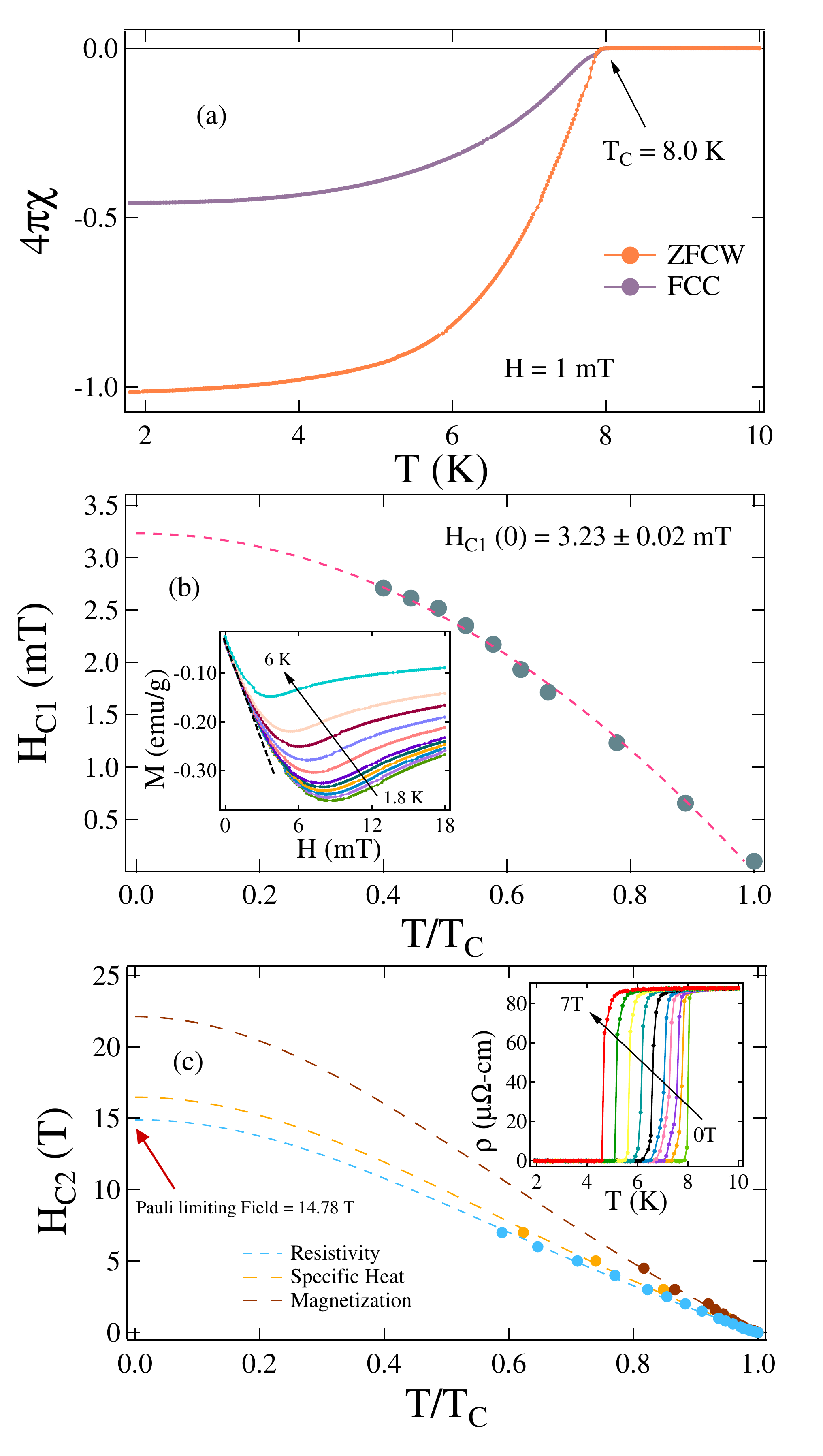}
\caption{ \label{Fig3:Mag} a) Temperature dependence of the magnetic susceptibility for Re$_{5.5}$Ta taken via ZFCW and FCC processes, shows T$_{C}$ at 8.0 K. b) Temperature dependence of H$_{C1}$ for Re$_{5.5}$Ta. Inset in \figref{Fig3:Mag}(b) represents low-field magnetization curves at various temperatures (1.8 $\leq$ T $\leq$ 6 K). c) Determination of upper critical field via specific heat, resistivity, and magnetization data where dotted lines represent fitting using \equref{eqn2:Hc2}. Inset displays resistivity variation with the temperature at different applied fields.}
\end{figure}

Figure \ref{Fig3:Mag}(a) shows the temperature variation of the magnetic susceptibility collected via zero-field cooled warming (ZFCW), and field cooled cooling (FCC) modes in an applied field of 1mT. Both regimes exhibit a diamagnetic signal at a transition temperature, T$_C$ = 8.0 $\pm$ 0.1 K.  A weak diamagnetic signal in the FCC data on entering the superconducting state is due to the magnetic flux pinning.

To determine the lower critical field $H_{C1}$, the low-field magnetization curves as a function of an applied magnetic field (0 - 18 mT) were measured at various temperatures and are shown in the inset of \figref{Fig3:Mag}(b). $H_{C1}$ is defined as the point deviating from the linear or the Meissner line for the initial slope in the magnetization curves. For each of the M-H curves, the point of deviation from the Meissner line is computed and analyzed using the Ginzburg-Landau relation (see main \figref{Fig3:Mag}(b)).
\\
\begin{equation}
H_{C1}(T)=H_{C1}(0)\left[1-\left(\frac{T}{T_{C}}\right)^{2}\right]
\label{eqn1:Hc1}
\end{equation} 
\\
$H_{C1}$(0) is estimated to be 3.23 $ \pm $ 0.02 mT. In order to calculate the upper critical field, H$_{C2}$(0), the effect of the applied magnetic field on T$_{C}$ is measured by various techniques, such as: resistivity, magnetization, and specific heat. The inset of \figref{Fig3:Mag}(c) represents the temperature variation of resistivity in the field range of 0 T to 7 T. An increase in the applied magnetic field suppresses T$_{C}$. In the magnetization measurement, the onset of a diamagnetic signal is considered as the transition temperature whereas, in the case of resistivity and specific heat data, the midpoint is taken as the criteria for T$_{C}$. Figure \ref{Fig3:Mag}(c) shows a linear response in H$_{C2}$(T) from all the above-mentioned measurements when plotted against reduced temperature t = T/T$_C$. As seen in the inset of \figref{Fig3:Mag}(c) that superconductivity is not suppressed even at 7 T, evidences a more significant value of the upper critical field. The temperature variation of H$_{C2}$(T) can be well described by the relation  
\\
\begin{equation}
H_{C2}(T) = H_{C2}(0)\left[\frac{(1-t^{2})}{(1+t^2)}\right]. 
\label{eqn2:Hc2}
\end{equation} 
\\
The data fits well in all the cases, yielding upper critical field, H$_{C2}$(0): 22.11 $\pm$ 0.32 T, 14.89 $\pm$ 0.13 T and 16.47 $\pm$ 0.43 T from magnetization, resistivity, and specific heat measurements respectively. 

According to BCS theory, the Pauli limiting field is given by $ H_{C2}^{P}(0)$ = \textit{const.}T$_{C} $ where \textit{const.} = 1.86 T/K \cite{Pauli_1,Pauli_2}. For Re$_{5.5}$Ta, $ H_{C2}^{P}(0)$ = 14.78 T, which is interestingly less than the H$_{C2}$(0) calculated from all the measurements. The orbital limit for an upper critical field is given by Werthamer-Helfand-Hohenberg (WHH) expression \cite{WHH_1,WHH_2} 

\begin{equation}
H_{C2}^{orbital}(0) = -\alpha T_{C}\left.\frac{dH_{C2}(T)}{dT}\right|_{T=T_{C}}
\label{eqn3:HHH}
\end{equation}
\\
The initial slope $\frac{-dH_{C2}(T)}{dT} $ at T = T$_{C}$ is estimated to be 1.69 $ \pm $ 0.24 T/K. Considering $\alpha$ = 0.693 gives the orbital limiting upper critical field $ H_{C2}^{orbital}(0)$ = 9.33 $\pm$ 0.79 T. The Maki parameter \cite{Maki} which is a measure of the relative strength of Pauli and orbital limits for the upper critical field, is derived from the expression $\alpha_{M} = \sqrt{2}H_{C2}^{orb}(0)/H_{C2}^{P}(0)$, yields $\alpha_{M} = 0.89 $. This value indicates the non-negligible effect of the Pauli limiting field for Re$_{5.5}$Ta \cite{Maki_Par}. The upper critical field H$_{C2}$(0) determined for Re$_{5.5}$Ta is higher than both the Pauli and orbital limiting fields, suggesting the probable presence of some triplet component in the superconducting ground state like in other Re$_{6}$X compounds \cite{Zr_TRS,FerLiqd_3&Re6Zr_WTRS}. 

The calculated values of H$_{C2}$(0) and H$_{C1}$(0) have been used to estimate two fundamental length scales of a superconductor: Ginzburg-Landau coherence length ($\xi_{GL}$(0)) is given by the following expression: $H_{C2}(0) = \frac{\Phi_{0}}{2\pi\xi_{GL}^{2}}$ \cite{Coh_Leng} where $ \Phi_{0}$  ( = 2.07 $\times$ 10$^{-15}$ Tm$^{2}$) is the magnetic flux quantum. Using $  H_{C2}(0) $ = 16.47 T, it is evaluated to be 44.83 $\pm$ 0.58 \text{\AA}. The obtained value of H$_{C1}$(0) and $\xi_{GL}$(0) provides information about Ginzburg-Landau penetration depth $\lambda_{GL}$(0) using the relation \cite{Pen_Dep}
\\
\begin{equation}
H_{C1}(0) = \frac{\Phi_{0}}{4\pi\lambda_{GL}^2(0)}\left(\mathrm{ln}\frac{\lambda_{GL}(0)}{\xi_{GL}(0)}+0.12\right)   
\label{eqn4:PD}
\end{equation} 
\\

\begin{figure} 
\includegraphics[width=1.0\columnwidth, origin=b]{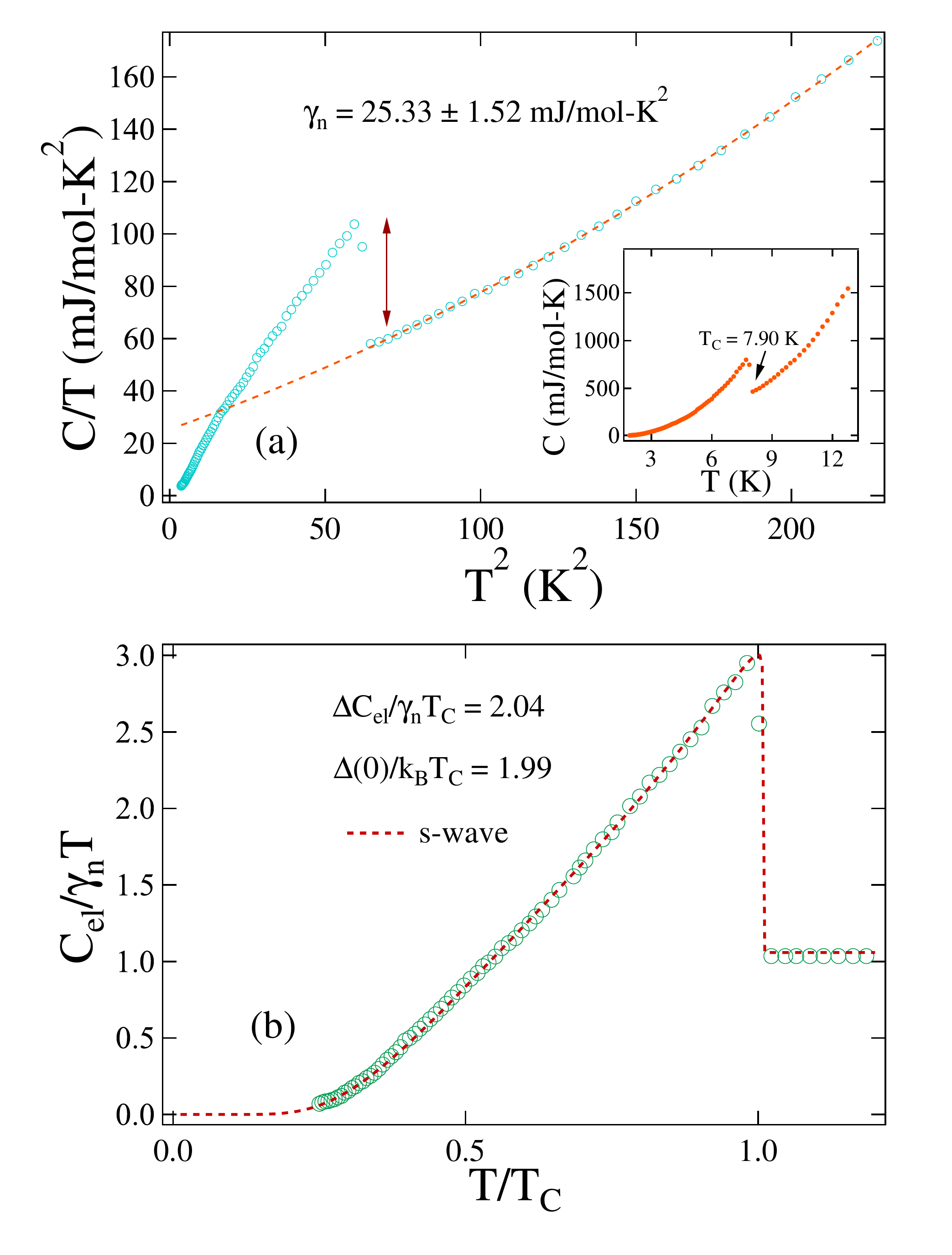}
\caption{ \label{Fig4:SH} a) C/T vs T$ ^{2}$ data fits well with \equref{eqn5:SH1} represented by dotted line. Inset: Temperature dependence of specific heat in zero-field exhibiting T$_{C}$ at 7.90 K. b) Normalized specific heat data, $C_{el}/\gamma_{n}T$, is well described by single gap s-wave model shown by dotted red line.}
\end{figure}

For $  H_{C1}(0) $ = 3.23 $\pm$ 0.02 mT  and $\xi_{GL}$(0) = 45 $\pm$ 1 \text{\AA}, $\lambda_{GL}$(0) is found out to be 4949 $\pm$ 9\text{\AA}. The Ginzburg-Landau parameter $\kappa_{GL}$ is related to $\xi_{GL}$(0) and $\lambda_{GL}$(0) by the relation: $\kappa_{GL} = \frac{\lambda_{GL}(0)}{\xi_{GL}(0)}$. Using the calculated values of $\xi_{GL}$(0) = 45 $\pm$ 1 \text{\AA} and $\lambda_{GL}$(0) = 4949 $\pm$ 9\text{\AA}, the value obtained for $\kappa_{GL}$ = 111 $\pm$ 1. It indicates that Re$_{5.5}$Ta is a strong type II superconductor. An estimation of the thermodynamic critical field H$_{C}$ has been done by using the relation \cite{Pen_Dep}: $H_{C1}(0)H_{C2}(0) = H_{C}^2\mathrm{ln}\kappa_{GL}$, and provides the value of H$ _{C} $ = 107 $\pm$ 3 mT.

\subsubsection{Specific heat}
Figure \ref{Fig4:SH}(a) represents the zero-field specific heat data. The discontinuity at the superconducting transition temperature, T$_{c}$ = 7.90 $\pm$ 0.10 K is in good agreement with the resistivity and magnetization measurements [see inset \figref{Fig4:SH}(a)]. Low-temperature specific heat data in \figref{Fig4:SH}(a) are well described by the relation:
\\\begin{equation}  
\frac{C}{T}=\gamma_{n}+\beta_{3} T^{2} + \beta_{5}T^{4}
\label{eqn5:SH1}    
\end{equation} 

where Sommerfeld coefficient, $\gamma_{n}$ represents electronic contribution, Debye constant, $ \beta_{3} $ provides the information regarding phononic contribution, and $ \beta_{5}$ represents the anharmonic contribution to the specific heat. Fitting  C/T Vs T$ ^{2}$ data with \equref{eqn5:SH1} gives: $ \gamma_{n}$ = 25.33 $\pm$ 1.52  mJmol$^{-1}$K$^{-2}$, $\beta_{3} = 0.42 \pm 0.02$  mJmol$^{-1}$K$^{-4}$ and $ \beta_{5} $ = 1.02 $ \pm $ 0.06 $\mu$Jmol$^{-1}$K$^{-6}$.

The Sommerfeld coefficient, $ \gamma_{n} $ is directly related to the density of states, $D_{C}(E_{\mathrm{F}})$ via the relation: $\gamma_{n}= \left(\frac{\pi^{2}k_{B}^{2}}{3}\right)D_{C}(E_{\mathrm{F}})$, where k$_{B}$ = 1.38$\times$10$^{-23}$ J K$^{-1}$. $D_{C}(E_{\mathrm{F}})$ is estimated to be 6.72 states eV$^{-1}$f.u$^{-1}$. From Debye constant $\beta_{3}$, Debye temperature $\theta_{D}$ is estimated which is related through the formula $\theta_{D}= \left(\frac{12\pi^{4}RN}{5\beta_{3}}\right)^{\frac{1}{3}}$, where R = 8.314 J mol$ ^{-1} $K$ ^{-1} $ is a gas constant, and N is the number of atoms per formula unit. The calculated value of $\theta_{D}$ = 310 $\pm$ 4 K. 
 
The $ \lambda_{e-ph} $ represents the electron-phonon coupling constant which gives the strength of attractive interaction between electron and phonon. In the McMillan's Model \cite{McMillan}, $ \lambda_{e-ph} $ takes into account the computed values of $\theta_{D}$ and T$_{C}$ as follows:
\\
\begin{equation}
\lambda_{e-ph} = \frac{1.04+\mu^{*}\mathrm{ln}(\theta_{D}/1.45T_{C})}{(1-0.62\mu^{*})\mathrm{ln}(\theta_{D}/1.45T_{C})-1.04 }
\label{eqn6:Lambda}
\end{equation}
\\
Here $\mu^{*}$ accounts for  screened Coulomb repulsion and is taken to be 0.13. By considering $\theta_{D}$ = 310 K and T$_{C}$ = 7.90 K, we obtain $ \lambda_{e-ph} $ = 0.73. This value of $ \lambda_{e-ph} $ classify Re$_{5.5}$Ta as a strongly coupled superconductor when compared with other moderately coupled NC superconductors such as Re$_{6}$Hf ($ \lambda_{e-ph} $ = 0.63) \cite{R6H_WTRS}, Re$_{24}$Ti$_{5}$ \cite{R24T5_WTRS} and Re$_{3}$Ta \cite{Fer_Liqd_1&N0.18R0.82} ($ \lambda_{e-ph} $ = 0.6).

Electronic contribution to specific heat, C$_{el}$, can be calculated by subtracting the phononic contribution, C$_{ph}  = \beta_{3} T^{3} + \beta_{5}T^{5}$, from total specific heat. The normalized magnitude of electronic specific heat jump $\frac{\Delta C_{el}}{\gamma_{n}T_{C}}$ = 2.04, which is significantly higher than the BCS value 1.43 in the weak coupling limit, suggesting that the superconducting state in Re$_{5.5}$Ta has a stronger electron-phonon coupling than the BCS superconductors \cite{Y5R6S18,K2C3A3}.

The low-temperature specific heat data was fitted by the single-gap BCS expression for normalized entropy S given by the following relation
\\
\begin{equation}
\frac{S}{\gamma_{n}T_{C}} = -\frac{6}{\pi^2}\left(\frac{\Delta(0)}{k_{B}T_{C}}\right)\int_{0}^{\infty}[ \textit{f}\ln(f)+(1-f)\ln(1-f)]dy \\
\label{eqn7:BCS1}
\end{equation}
\\
where  $\textit{f}$($\xi$) = [exp($\textit{E}$($\xi$)/$k_{B}T$)+1]$^{-1}$ is the Fermi function, $\textit{E}$($\xi$) = $\sqrt{\xi^{2}+\Delta^{2}(t)}$, where E($ \xi $) is the energy of the normal electrons measured relative to Fermi energy, $\textit{y}$ = $\xi/\Delta(0)$, $\mathit{t = T/T_{C}}$ and $\Delta(t)$ = tanh[1.82(1.018(($\mathit{1/t}$)-1))$^{0.51}$] is the BCS approximation for the temperature dependence of energy gap. The normalized entropy S is related to the normalized electronic specific heat by 
\\
\begin{equation}
\frac{C_{el}}{\gamma_{n}T_{C}} = t\frac{d(S/\gamma_{n}T_{C})}{dt} \\
\label{eqn8:BCS2}
\end{equation}

The dotted line in \figref{Fig4:SH}(b) is the fit to the data using single gap BCS model. From this fit we obtain, $\alpha$ = $\Delta(0)/k_{B}T_{C}$ = 1.99, which is higher than the BCS value of $ \alpha $ = 1.764, indicating a strongly coupled superconductivity. 

\subsubsection{Muon spin rotation and relaxation}

\begin{figure}
\includegraphics[width=1.0\columnwidth]{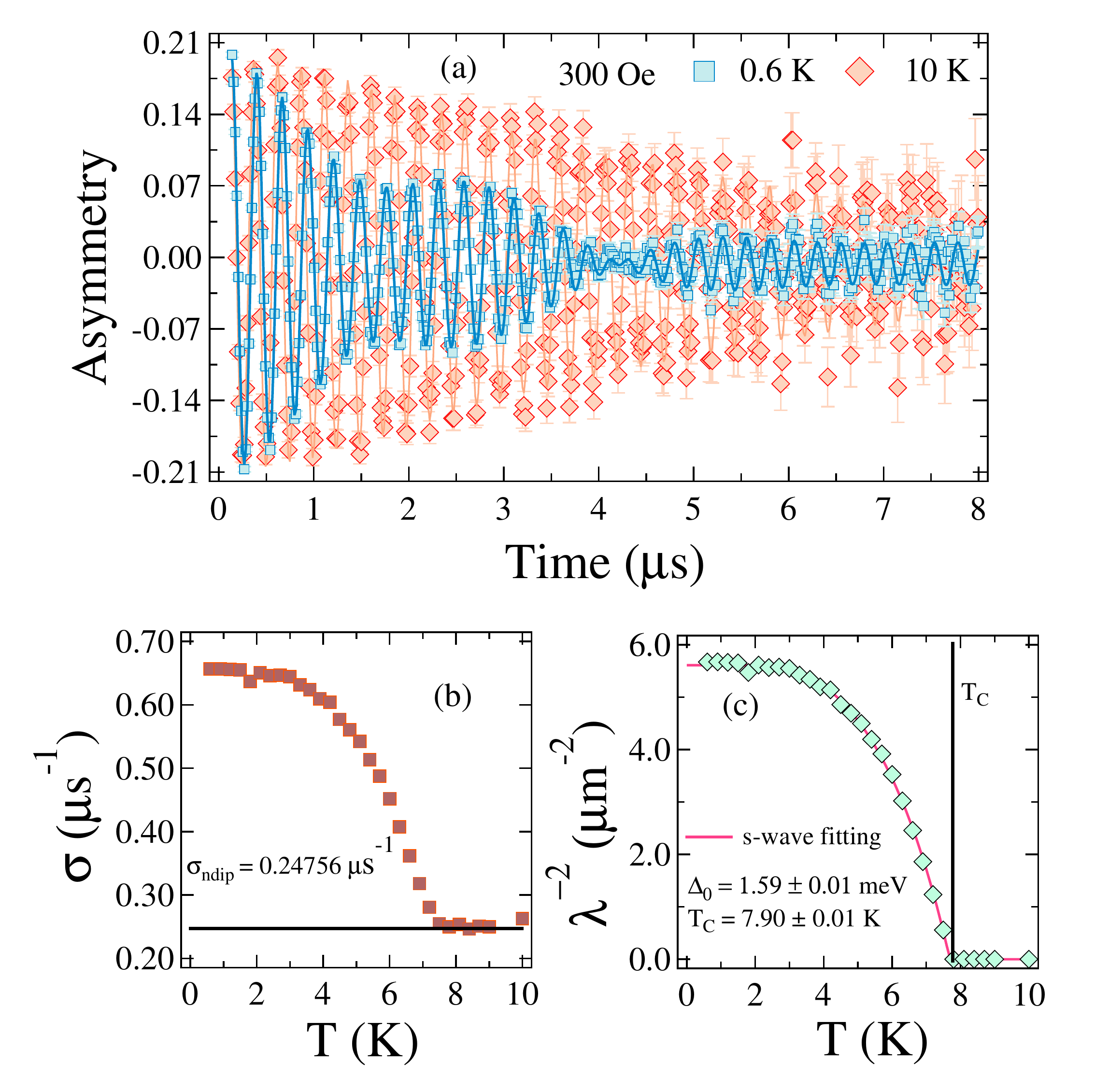}
\caption{\label{Fig5:TF} a) Transverse field spectra recorded in an applied magnetic field of 30 mT at temperature 10 K ( >  $T_{c}$)  and 0.6 K (<  $T_{c}$). The solid lines are fits using \equref{eqn9:TF1}. b) Temperature dependence of total depolarization rate, $\sigma$, using second moment method. c) Variation of $\lambda^{-2}$ with respect to temperature for Re$_{5.5}$Ta where line represents s-wave fitting using \equref{eqn14:swave}}.
\end{figure}

To investigate the superconducting ground state of Re$_{5.5}$Ta at a microscopic level, muon spin rotation and relaxation measurements were carried out. TF $\mu$SR measurement was performed in an applied field of 30 mT perpendicular to the initial muon spin polarization and above H$_{C1}$ and below H$_{C2}$, thus allowing the generation of a flux line lattice in the mixed superconducting state. Typical asymmetry spectra recorded above and below T$_{C}$ are displayed in \figref{Fig5:TF}(a). The time-domain spectra were best fitted by the decaying Gaussian oscillatory function given below

\begin{equation}
\begin{split}
A (t) = \sum_{i=1}^N A_{i}\exp\left(-\frac{1}{2}\sigma_i^2t^2\right)\cos(\gamma_\mu B_it+\phi)\\ + A_{bg}\cos(\gamma_\mu B_{bg}t+\phi),
\label{eqn9:TF1}
\end{split}
\end{equation}

where $\phi$, $A_{i}$, $B_{i}$, $\sigma_{i}$ and $\gamma_{\mu}/2\pi$ = 135.5 MHz/T  are the initial phase, asymmetry, mean field of the ith component of the Gaussian distribution, relaxation rate and muon gyromagnetic ratio, respectively. $A_{bg}$ and $B_{bg}$ are the background contributions for the asymmetry and the field which originates from the muon hitting the sample holder or the walls of the cryostat. We found that two Gaussian components (N=2) are sufficient to fit the asymmetry spectra for our sample. The second moment method has been used to calculate the total depolarization rate $\sigma$ and its temperature dependence, as displayed in \figref{Fig5:TF}(b). The first and second moments are described as:

\begin{equation}
\expval{B} = \sum_{i=1}^2\frac{A_{i}B_{i}}{A_{1}+A_{2}},
\label{eqn10:TF2}
\end{equation}

\begin{equation}
\expval{\Delta B^2} = \frac{\sigma^2}{\gamma_{\mu}^2} = \sum_{i=1}^2\frac{A_{i}[(\sigma_{i}/\gamma_{\mu})^2+(B_{i}-\expval{B})^2]}{A_{1}+A_{2}},
\label{eqn11:TF3}
\end{equation}

Above the transition temperature T$_{C}$, $\sigma$ has an almost constant value which is attributed to the nuclear dipolar field, $\sigma_{ndip}$. To extract the superconducting contribution $\sigma_{sc}$, $\sigma_{ndip}$ = 0.24756 was subtracted from $\sigma$ using the following expression
\begin{equation}
\sigma_{\mathrm{sc}} = \sqrt{\sigma^{2} - \sigma_{\mathrm{ndip}}^{2}}
\label{eqn12:sigma}
\end{equation}
For a superconductor with hexagonal Abrikosov flux line lattice and large upper critical field, the penetration depth, $\lambda$, is related to $\sigma_{sc}$ as follows \cite{SigSC_1, SigSC_2}:
\begin{equation}
\frac{\sigma_{\mathrm{sc}}^2(T)}{\gamma_{\mu}^2} = \frac{0.00371\Phi_{0}^2}{\lambda^{4}(T)}
\label{eqn13:sigmaH}
\end{equation}
where $\gamma_{\mu}/2\pi$ = 135.5 MHz/T is the muon gyromagnetic ratio, and $\Phi_{0}$ is the magnetic flux quantum. The temperature dependence of calculated $\lambda^{-2}$ is presented in \figref{Fig5:TF}(c). At low temperatures (T$_{C}$/3), $\lambda^{-2}$ appears to be flattened or temperature independent. This indicates fully gapped or s-wave superconductivity and excludes the possibility of nodes in the energy gap. This also yields the magnetic penetration depth at T = 0 K, $\lambda(0)$ = 4226 $\pm$ 9 $\text{\AA}$ which is slightly less than the value found from magnetization measurements.

For further verification of the nature of superconducting gap, the temperature dependence of the London penetration depth $\lambda(T)$ for a s-wave BCS superconductor in the dirty limit within the local London approximation has been studied using the following expression:
\begin{equation}
\frac{\lambda^{-2}(T)}{\lambda^{-2}(0)} = \frac{\Delta(T)}{\Delta(0)}\mathrm{tanh}\left[\frac{\Delta(T)}{2k_{B}T}\right] ,
\label{eqn14:swave}
\end{equation}
where  $ \Delta $(T) = $ \Delta_{0} $ tanh[1.82(1.018($\mathit{(T_{C}/T})$-1))$^{0.51}$] is the BCS approximation for the temperature dependence of the energy gap. This fit yields a value of energy gap $\Delta(0)$ = 1.59 meV and $\Delta(0)/k_{B}T_{C}$ = 2.39 which is much higher than the expected BCS value 1.76. It is indicative of strong electron-phonon coupling in the superconducting state of Re$_{5.5}$Ta consistent with the specific heat data. Such an enhanced value of the superconducting gap is also observed in other strongly coupled NC superconductors \cite{L5R6S18,Y5R6S18,K2C3A3,ThCoC2,BP}.

\begin{figure}
\includegraphics[width=1.0\columnwidth]{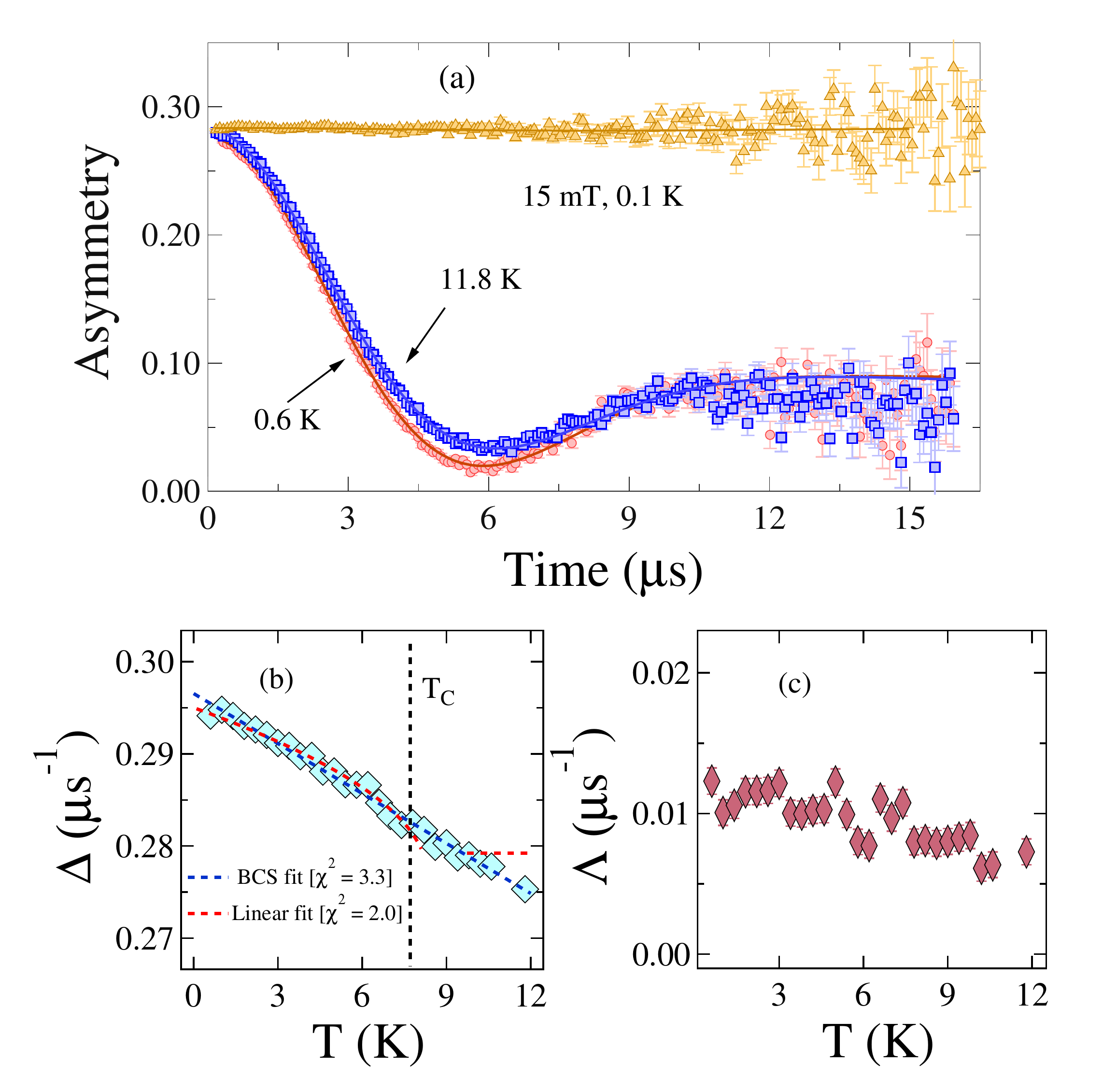}
\caption{\label{Fig6:ZF} a) Zero-field asymmetry spectra collected below (T = 0.6 K) and above (T = 11.8 K) the transition temperature, T$_{C}$ where solid lines represents fit to the data using \equref{eqn16:ZF2}. Longitudinal spectrum is also shown by orange triangle recorded at 0.1 K and in an applied field of 15 mT. b) Temperature variation of the relaxation rate $\Delta$ indicating spontaneous fields appearing in the superconducting state of Re$_{5.5}$Ta. c) Variation of electronic relaxation rate $\Lambda$ with respect to temperature. }
\end{figure}

In order to address the question regarding the time-reversal symmetry breaking in Re$_{5.5}$Ta, zero-field muon spin relaxation (ZF-$\mu$SR) measurements were carried out. The time-domain relaxation spectra are recorded below ($T$ = 0.6 K) and above ($T$ = 11.8 K) the transition temperature, $T_{c}$ = 8.0 K and are shown in \figref{Fig6:ZF}(a). The absence of oscillatory components in the spectra rules out the presence of any ordered magnetic structure. There is an appreciable change in the relaxation rates of the spectra recorded below and above the transition temperature $T_{c}$. The relaxation rate is stronger in the superconducting state in comparison to the normal state, which indicates the presence of internal magnetic fields in the superconducting state. 

In ZF-$\mu$SR, where at low-temperatures, muon diffusion is not appreciable, the behaviour of asymmetry spectra are modelled by the Gaussian Kubo-Toyabe (KT) function \cite{KT} 

\begin{equation}
G_{\mathrm{KT}}(t) = \frac{1}{3}+\frac{2}{3}(1-\Delta^{2}t^{2})\mathrm{exp}\left(\frac{-\Delta^{2}t^{2}}{2}\right) ,
\label{eqn15:ZF}
\end{equation} 
where $\Delta$ represents the muon spin relaxation due to the randomly oriented, static nuclear moments experienced at the muon site. The best description of the spectra in zero-field is given by the following function:
\begin{equation}
A(t) = A_{0}G_{\mathrm{KT}}(t)\mathrm{exp}(-\Lambda t)+A_{1} ,
\label{eqn16:ZF2}
\end{equation}
where $A_{0}$ is the initial sample related asymmetry, $A_{1}$ is background contribution to asymmetry from the muons stopping in the sample holder whereas $\Lambda$ accounts for the electronic relaxation rates.

The solid lines in \figref{Fig6:ZF}(a) represent the fits, using \equref{eqn16:ZF2}, and yields the temperature dependence of $\Delta$(T). The shape of $\Delta$(T) gives rise to several possibilities: the significant increase of $\Delta$ at low temperatures suggests a spontaneous emergence of the internal magnetic field in the superconducting state of Re$_{5.5}$Ta and point towards a possible TRSB signal. The change in relaxation rate is quite evident, with an increase of 0.015 $\mu$s$^{-1}$ below T$_{C}$, which corresponds to a characteristic field strength $|B_{int}|$ = $\sqrt2\Delta/\gamma_{\mu}$ = 0.25 G. Interestingly, a similar pattern has been exhibited by $\Delta$(T) in Nb$_{0.18}$Re$_{0.82}$ \cite{Re-Nb} which ascertains TRSB in the compound. In fact, a small longitudinal field of 15 mT was sufficient to decouple the muon spins from the internal magnetic field as shown in \figref{Fig6:ZF}(a), indicating that the relevant fields in the spectra are static or quasistatic on the time scale of the muon precession. Despite the clear difference in the ZF-$\mu$SR spectra recorded in the normal and superconducting states [see \figref{Fig6:ZF}(a)], $\Delta$(T) does not show distinct changes across T$_{C}$. In order to check how the increase of $\Delta$(T) is correlated with T$_{C}$, two types of fit have been made using (a) BCS-type function where T$_{C}$ = 8 K was kept fixed and (b) linear function which takes into account the whole temperature range. The goodness of fit is better for the linear function than the BCS-type function which point towards the probable presence of spin fluctuations, proposed in a few compounds such as Cs$_{2}$Cr$_{3}$As$_{3}$ \cite{CsCrAs} and RRuB$_{2}$ (R=Lu,Y) \cite{RRuB}. Though the later proposed behaviour has been observed only in $\Lambda$ for the mentioned compounds, an analogy has been drawn in the case of Re$_{5.5}$Ta. A similar type of behaviour in $\Delta$(T) has also been observed in a compound, Be$_{22}$Re \cite{B22R}. The electronic relaxation rate $\Lambda(T)$ remains mostly constant in the studied temperature range as shown in \figref{Fig6:ZF}(c). Therefore, from the present results obtained via ZF-$\mu$SR measurement, a definite conclusion cannot be drawn regarding a clear signature for TRSB in Re$_{5.5}$Ta. The possibility of spin fluctuations in $\Delta$(T) can be due to the high atomic percentage of Re in Re$_{5.5}$Ta. In order to see how the spin fluctuations and TRSB signal correlated with Re atomic percentage, further experiments such as muon spin relaxation, NMR on a Re-Ta compound containing a less atomic percentage of Re must be performed. Moreover, the presence of spin fluctuations in the system can be detected by NMR and NQR measurements.

\subsubsection{Electronic properties and the Uemura plot}

In order to determine the London penetration depth, electronic mean free path and to verify dirty limit superconductivity for Re$_{5.5}$Ta, a following set of equations has been used. The relation between Sommerfeld coefficient, quasiparticle number density per unit volume and mean free path is given via the expression
\begin{equation}
\gamma_{n} = \left(\frac{\pi}{3}\right)^{2/3}\frac{k_{B}^{2}m^{*}V_{\mathrm{f.u.}}n^{1/3}}{\hbar^{2}N_{A}}
\label{eqn17:gf}
\end{equation}
where k$_{B}$ is the Boltzmann constant, V$_{\mathrm{f.u.}}$ is the volume of a formula unit, N$_{A}$ is the Avogadro number and m$^{*}$ is the effective mass of quasiparticles. Residual resistivity is related to Fermi velocity $v_{\mathrm{F}}$, and electronic mean free path $\textit{l}$ by the expression
 \begin{equation}
\textit{l} = \frac{3\pi^{2}{\hbar}^{3}}{e^{2}\rho_{0}m^{*2}v_{\mathrm{F}}^{2}}
\label{eqn18:le}
\end{equation}
whereas the Fermi velocity $v_{\mathrm{F}}$ can be expressed in terms of quasiparticle carrier density and effective mass by
\begin{equation}
n = \frac{1}{3\pi^{2}}\left(\frac{m^{*}v_{\mathrm{f}}}{\hbar}\right)^{3} .
\label{eqn19:n}
\end{equation}
The expression for the penetration depth $\lambda_{GL}$(0) in dirty limit is given by
\begin{equation}
\lambda_{GL}(0) = \left(\frac{m^{*}}{\mu_{0}n e^{2}}\right)^{1/2}\left(1+\frac{\xi_{0}}{\textit{l}}\right)^{1/2}
\label{eqn20:f}
\end{equation}
where $\xi_{0}$ is the BCS coherence length and the first term in the bracket represents the London penetration depth $\lambda_{L}$. In the dirty limit at T = 0 K, Ginzburg-Landau coherence length $\xi_{GL}$(0) and BCS coherence length $\xi_{0}$ are related by the expression
\begin{equation}
\frac{\xi_{GL}(0)}{\xi_{0}} = \frac{\pi}{2\sqrt{3}}\left(1+\frac{\xi_{0}}{\textit{l}}\right)^{-1/2}
\label{eqn21:xil}
\end{equation}

\begin{figure}
\includegraphics[width=1.0\columnwidth]{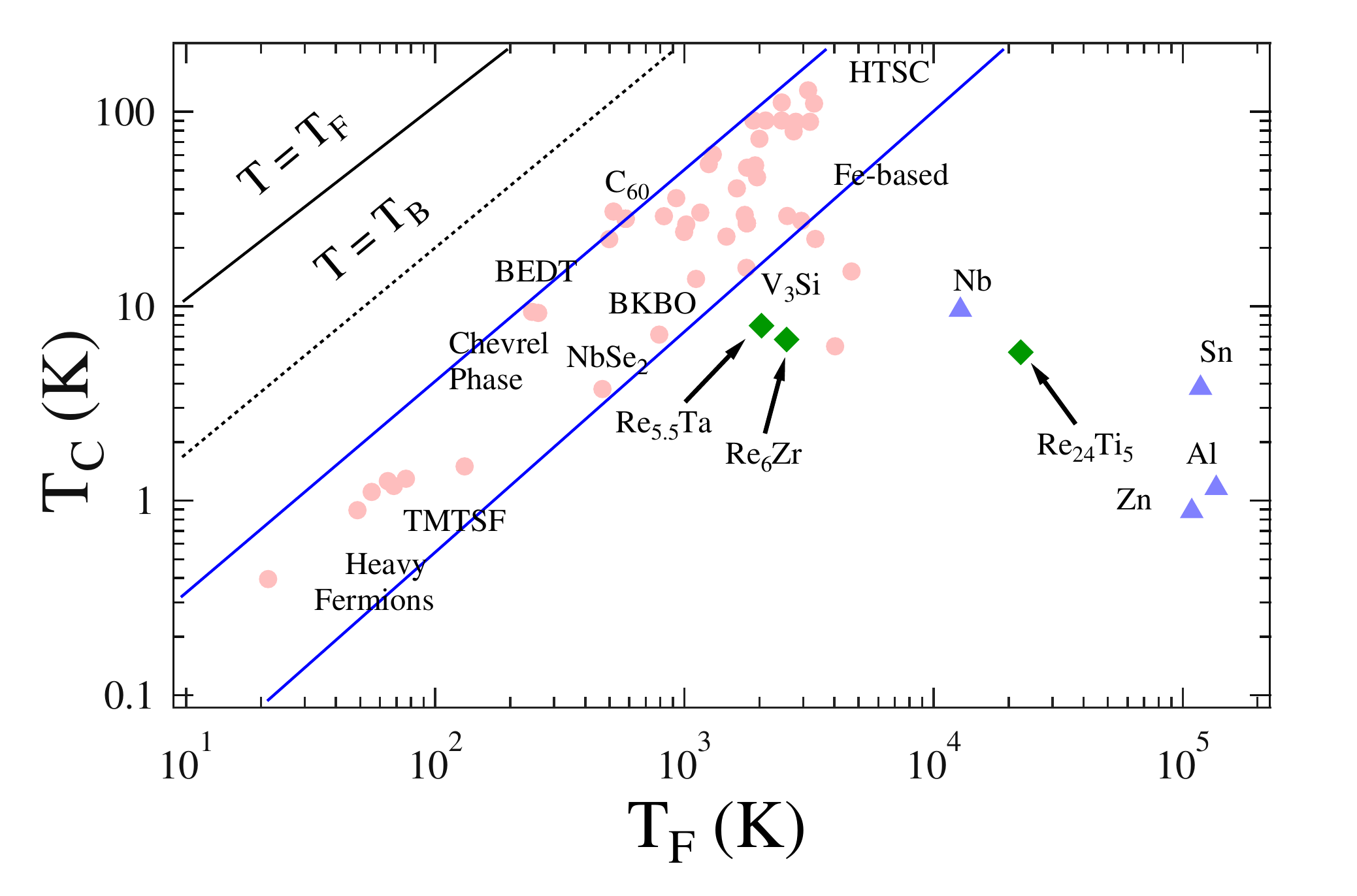}
\caption{\label{Fig7:UP} A plot between the superconducting transition temperature $T_{c}$ and the effective Fermi temperature $T_{F}$ where Re$_{5.5}$Ta along with some other members of $\alpha$-Mn family is shown by solid green markers \cite{Umera_Ref,R24T5_WTRS}. The data points \cite{Unconv_1,Unconv_2} between two solid blue lines represent the band of unconventionality.} 
\end{figure}

\begin{table}[h!]
\caption{Parameters in the superconducting and normal state of Re$_{5.5}$Ta and Re$_{3}$Ta \cite{R3T}}
\begingroup
\setlength{\tabcolsep}{12pt}
\begin{tabular}{c c c c} 
\hline\hline
Parameters & Unit & Re$_{5.5}$Ta & Re$_{3}$Ta\\ [1ex]
\hline
$T_{C}$& K& 8.0& 4.68\\             
$H_{C1}(0)$& mT& 3.23& 2.13\\                       
$H_{C2}(0)$& T& 16.4& 7.3\\
$H_{C2}^{P}(0)$& T& 14.78& 9.08\\
$H_{C2}^{Orb}(0)$& T& 9.33& 8.4\\
$\xi_{GL}$& \text{\AA}& 45& 67.1\\
$\lambda_{GL}$& \text{\AA}& 4949& 5880\\
$k_{GL}$& &111& 88\\
$\gamma_{n}$&  mJ mol$^{-1}$ K$^{-2}$& 25.3& 13.1\\                      
$\theta_{D}$& K& 310& 321\\
$\Delta C_{el}/\gamma_{n}T_{C}$&   &2.04& 1.51\\
$\Delta(0)/k_{B}T_{C}$&  &1.99& 1.84\\ 
$\xi_{0}/l_{e}$&   &5& 0.86\\
$v_{F}$& m s$^{-1}$& 121770& 37000\\
$n$& 10$^{27}$m$^{-3}$& 2.8& 3.3\\
$T_{F}$& K& 2040& 640\\
$T_{C}/T_{F}$& & 0.0038& 0.0073\\
$m^{*}$/m$_{e}$&  & 4.15& 14.5\\
[1ex]
\hline\hline
\end{tabular}
\endgroup
\end{table}

The above Eqs. (\ref{eqn17:gf}-\ref{eqn21:xil}) were solved simultaneously to estimate the parameters m$^{*}$, n, $\textit{l}$, and $\xi_{0}$ as done in Ref. \cite{Umera_Ref} by using the values of $\gamma_{n}$ = 25.33 $ \pm $ 1.52 mJ mol$^{-1}$K$^{-2}$, $\xi_{GL}$(0) = 45 $\pm$ 1 \text{\AA}, and $\rho_{0}$ = 89.95 $\pm$ 0.08 $\mu$ $\Omega$-cm. All the calculated parameters are listed in Table II, and the ratio of $\xi_{0}$/$\textit{l}$ clearly indicates that Re$_{5.5}$Ta is in the dirty limit. The order of calculated parameters matches well with other non-centrosymmetric $\alpha$-Mn superconductors where dirty limit superconductivity was observed \cite{R3T,Umera_Ref}.

Uemura et al. \cite{Umera} provide a classification between conventional and unconventional superconductors according to their $\frac{T_{C}}{T_{F}}$ ratio. If this ratio falls in the range 0.01 $\leq$ $\frac{T_{C}}{T_{F}}$ $\leq$ 0.1, then that particular material is classified as unconventional superconductor. In order to classify Re$_{5.5}$Ta, the Fermi temperature ${T_{F}}$ \cite{Tf} has been calculated by the following expression
\begin{equation}
 k_{B}T_{F} = \frac{\hbar^{2}}{2}(3\pi^{2})^{2/3}\frac{n^{2/3}}{m^{*}}, 
\label{eqn13:tf}
\end{equation}
where n is the quasi-particle number density per unit volume, and m$^{*}$ is the effective mass of quasi-particles. By considering both the calculated values of n and m$^{*}$ listed in Table II, the estimated value of $T_{F}$ = 2040 K for Re$_{5.5}$Ta. The ratio $\frac{T_{C}}{T_{F}}$ has been found out to be 0.0038 which places Re$_{5.5}$Ta closer to the band of unconventional superconductors side among all the other members of $\alpha$-Mn family \cite{Umera_Ref,R24T5_WTRS}, shown by solid green symbols in \figref{Fig7:UP} and the solid blue lines representing the band of unconventional superconductors.

\section{CONCLUSION}
Detailed transport, magnetization, specific heat and muon spectroscopy measurements have been carried out on Re$_{5.5}$Ta. The above measurements show type-II superconductivity with T$_{C}$ = 7.95 K, with strong electron-phonon coupling. Interestingly, the calculated value of H$_{C2}$(0) ($\sim$ 16.9 T) is higher than the Pauli paramagnetic limit H$_{P}$ ($\sim$ 14.7 T), which suggests the possible presence of mixed spin singlet and triplet superconducting ground state similar to other Re$_{6}$X series of compounds. A comparison between superconducting state parameters of Re$_{5.5}$Ta and Re$_{3}$Ta has been made in Table II. The enhanced superconducting transition temperature, upper critical field, and $\frac{\Delta C_{el}}{\gamma_{n}T_{C}}$ indicates that unconventional superconducting properties in the compounds of Re$_{6}$X series may arise because of the presence of high atomic \% of Re. It also suggests that the critical amount of high atomic number elements can be used to induce unconventional superconducting properties in the standard BCS superconductors. The temperature dependence of both electronic specific heat and TF-$\mu$SR data suggests a fully gapped s-wave superconductivity. ZF-$\mu$SR measurement suggests a probable presence of spin fluctuations and TRSB cannot be concluded in the superconducting state of Re$_{5.5}$Ta. To develop a clearer understanding of the nature of the superconducting ground state, it will be necessary to find the origin of the temperature dependence of $\Delta$ in ZF measurement and further experimental work is required on the single crystal of Re$_{5.5}$Ta.

\section{Acknowledgments}

R.~P.~S.\ acknowledges Science and Engineering Research Board, Government of India for the Core Research Grant CRG/2019/001028.


\begin{thebibliography}{0}%
\makeatletter
\providecommand \@ifxundefined [1]{%
 \@ifx{#1\undefined}
}%
\providecommand \@ifnum [1]{%
 \ifnum #1\expandafter \@firstoftwo
 \else \expandafter \@secondoftwo
 \fi
}%
\providecommand \@ifx [1]{%
 \ifx #1\expandafter \@firstoftwo
 \else \expandafter \@secondoftwo
 \fi
}%
\providecommand \natexlab [1]{#1}%
\providecommand \enquote  [1]{``#1''}%
\providecommand \bibnamefont  [1]{#1}%
\providecommand \bibfnamefont [1]{#1}%
\providecommand \citenamefont [1]{#1}%
\providecommand \href@noop [0]{\@secondoftwo}%
\providecommand \href [0]{\begingroup \@sanitize@url \@href}%
\providecommand \@href[1]{\@@startlink{#1}\@@href}%
\providecommand \@@href[1]{\endgroup#1\@@endlink}%
\providecommand \@sanitize@url [0]{\catcode `\\12\catcode `\$12\catcode
  `\&12\catcode `\#12\catcode `\^12\catcode `\_12\catcode `\%12\relax}%
\providecommand \@@startlink[1]{}%
\providecommand \@@endlink[0]{}%
\providecommand \url  [0]{\begingroup\@sanitize@url \@url }%
\providecommand \@url [1]{\endgroup\@href {#1}{\urlprefix }}%
\providecommand \urlprefix  [0]{URL }%
\providecommand \Eprint [0]{\href }%
\providecommand \doibase [0]{http://dx.doi.org/}%
\providecommand \selectlanguage [0]{\@gobble}%
\providecommand \bibinfo  [0]{\@secondoftwo}%
\providecommand \bibfield  [0]{\@secondoftwo}%
\providecommand \translation [1]{[#1]}%
\providecommand \BibitemOpen [0]{}%
\providecommand \bibitemStop [0]{}%
\providecommand \bibitemNoStop [0]{.\EOS\space}%
\providecommand \EOS [0]{\spacefactor3000\relax}%
\providecommand \BibitemShut  [1]{\csname bibitem#1\endcsname}%
\let\auto@bib@innerbib\@empty
\end{thebibliography}%


\begin{thebibliography}{References}

\bibitem{CPS} E. Bauer, G. Hilscher, H. Michor, Ch. Paul, E.W. Scheidt, A. Gribanov, Yu. Seropegin, H. Noel, M. Sigrist, and P. Rogl, Phys. Rev. Lett. 92, 027003 (2004).

\bibitem{CRS_1} N. Kimura, K. Ito, and H. Aoki, Phys. Rev. Lett. 95, 247004 (2005).

\bibitem{CRS_2} N. Kimura, K. Ito, H. Aoki, S. Uji, and T. Terashima, Phys. Rev. Lett. 98, 197001 (2007).

\bibitem{LP_PB} H. Q. Yuan, D. F. Agterberg, N. Hayashi, P. Badica, D. Vandervelde, K. Togano, M. Sigrist, and M. B. Salamon, Phys. Rev. Lett. 97, 017006 (2006).

\bibitem{KCA} Y. T. Shao, X. X. Wu, L. Wang, Y. G. Shi, J. P. Hu, J. L. Luo, EPL 123, 57001 (2018).

\bibitem{BP1} L. Jiao, J. L. Zhang, Y. Chen, Z. F. Weng, Y. M. Shao, J. Y. Feng, X. Lu, B. Joshi, A. Thamizhavel, S. Ramakrishnan, and H. Q. Yuan, Phys. Rev. B 89, 060507(R) (2014).

\bibitem{MIB} T. Klimczuk, F. Ronning, V. Sidorov, R. J. Cava, and J. D. Thompson, Phys. Rev. Lett. 99, 257004 (2007).

\bibitem{Sing-Trip_1} L. P. Gor'kov and E. I. Rashba, Phys. Rev. Lett. 87, 037004 (2001).
 
 \bibitem{Sing-Trip_2} S. Fujimoto, J. Phys. Soc. Jpn. 76, 051008 (2007).
 
 \bibitem{Pauli1} A. B. Karki, Y. M. Xiong, I. Vekhter, D. Browne, P. W. Adams, D. P. Young, K. R. Thomas, Julia Y. Chan, H. Kim, and R. Prozorov, Phys. Rev. B 82, 064512 (2010).
 
 \bibitem{Pauli2} Jin-Ke Bao, Ji-Yong Liu, Cong-Wei Ma, Zhi-Hao Meng, Zhang-Tu Tang, Yun-Lei Sun, Hui-Fei Zhai, Hao Jiang, Hua Bai, Chun-Mu Feng, Zhu-An Xu, and Guang-Han Cao, Phys. Rev. X 5, 011013 (2015).

 \bibitem{LPB_nod}  H. Takeya, M. ElMassalami, S. Kasahara, and K. Hirata, Phys. Rev. B 76, 104506 (2007).
 
 \bibitem{YC_nod} J. Chen, M. B. Salamon, S. Akutagawa, J. Akimitsu, J. Singleton, J. L. Zhang, L. Jiao, and H. Q. Yuan, Phys. Rev. B 83, 144529 (2011).
 
 \bibitem{LC_2gap}  S. Kuroiwa, Y. Saura, J. Akimitsu, M. Hiraishi, M. Miyazaki, K. H. Satoh, S. Takeshita, and R. Kadono, Phys. Rev. Lett. 100, 097002 (2008).
 
 \bibitem{LNC_2gap} J. Chen, L. Jiao, J. L. Zhang, Y. Chen, L. Yang, M. Nicklas, F. Steglich and H. Q. Yuan, New Journal of Physics 15, 053005 (2013).
 
 \bibitem{Zr_TRS} R. P. Singh, A. D. Hillier, B. Mazidian, J. Quintanilla, J. F. Annett, D. McK. Paul, G. Balakrishnan, and M. R. Lees, Phys. Rev. Lett. 112, 107002 (2014).
 
 \bibitem{Hf_TRS} D. Singh, J. A. T. Barker, A. Thamizhavel, D. McK. Paul, A. D. Hillier, and R. P. Singh, Phys. Rev. B 96, 180501(R) (2017).
 
 \bibitem{R6T} D. Singh, Sajilesh K. P., J. A. T. Barker, D. McK. Paul, A. D. Hillier, and R. P. Singh, Phys. Rev. B 97, 100505(R) (2018).
  
 \bibitem{R4.8T} T. Shang, G. M. Pang, C. Baines, W. B. Jiang, W. Xie, A. Wang, M. Medarde, E. Pomjakushina, M. Shi, J. Mesot, H. Q. Yuan, and T. Shiroka, Phys. Rev. B 97, 020502(R) (2018).
 
 \bibitem{Re-Nb} T. Shang, M. Smidman, S. K. Ghosh, C. Baines, L. J. Chang, D. J. Gawryluk, J. A. T. Barker, R. P. Singh, D. McK. Paul, G. Balakrishnan, E. Pomjakushina, M. Shi, M. Medarde, A. D. Hillier, H. Q. Yuan, J. Quintanilla, J. Mesot, and T. Shiroka, Phys. Rev. Lett. 121, 257002 (2018).

 \bibitem{LI} J.A.T. Barker, D. Singh, A. Thamizhavel, A.D. Hillier, M.R. Lees, G. Balakrishnan, D. McK. Paul, and R.P. Singh, Phys. Rev. Lett. 115, 267001 (2015).
 
  \bibitem{LNC_TRS} A. D. Hillier, J. Quintanilla, and R. Cywinski, Phys. Rev. Lett. 102, 117007 (2009).
 
  \bibitem{R3W_1} P. K. Biswas, A. D. Hillier, M. R. Lees, and D. McK. Paul, Phys. Rev. B 85, 134505 (2012).
  
   \bibitem{R3W_2} P. K. Biswas, M. R. Lees, A. D. Hillier, R. I. Smith, W. G. Marshall, and D. McK. Paul, Phys. Rev. B 94, 184592 (2011).
  
   \bibitem{R3T} J. A. T. Barker, B. D. Breen, R. Hanson, A. D. Hillier, M. R. Lees, G. Balakrishnan, D. McK. Paul, and R. P. Singh, Phys. Rev. B 98, 104506(R) (2018).
   
   \bibitem{Muon} MuonScience: Muons in Physics, Chemistry and Materials, edited by S. L. Lee, S. H. Kilcoyne, and R. Cywinski (Taylor and Francis, Abingdon, 1999).
   
   \bibitem{FerLiqd_3&Re6Zr_WTRS} Mojammel A. Khan, A. B. Karki, T. Samanta, D. Browne, S. Stadler, I. Vekhter, Abhishek Pandey, P. W. Adams, D. P. Young, S. Teknowijoyo, K. Cho, R. Prozorov, and D. E. Graf, Phys. Rev. B 94, 144515 (2016).
   
   \bibitem{R24T5_WTRS} C. S. Lue, H. F. Liu, C. N. Kuo, P. S. Shih, J. Y. Lin, Y. K. Kuo, M. W. Chu, T. L. Hung and Y. Y. Chen, Supercond. Sci. Technol. 26,  055011 (2013).
 
  \bibitem{Fer_Liqd_1&N0.18R0.82}  A. B. Karki, Y. M. Xiong, N. Haldolaarachchige, S. Stadler, I. Vekhter, P. W. Adams, D. P. Young, W. A. Phelan, and J. Y. Chan, Phys. Rev. B 83, 144525 (2011).
  
  \bibitem{Fer_Liqd_2} F. von Rohr, H. Luo, N. Ni, M. Worle, and R. J. Cava, Phys. Rev. B 89, 224504 (2014).
  
  \bibitem{KWR_1} K. Kadowaki and S. B. Woods, Solid State Commun. 58, 507 (1986).

 \bibitem{KWR_2} R. Jin, J. He, S. McCall, C. S. Alexander, F. Drymiotis, and D. Mandrus, Phys. Rev. B 64, 180503 (2001).

 \bibitem{KWR_3} A. C. Jacko, J. O. Fjaerestad, and B. J. Powell, Nat. Phys. 5, 422 (2009).
   
   \bibitem{Pauli_1} B. S. Chandrasekhar, Appl. Phys. Lett. 1, 7 (1962).
  
  \bibitem{Pauli_2} A.	M. Clogston, Phys. Rev. Lett. 9, 266 (1962).
  
  \bibitem{WHH_1} E. Helfand, and N. R. Werthamer, Phys. Rev. 147, 288 (1966)
  
  \bibitem{WHH_2} N. R. Werthamer, E. Helfand, and P. C. Hohenberg, Phys. Rev. 147, 295 (1966).
  
   \bibitem{Maki} K. Maki, Phys. Rev. B 148, 362 (1966).
  
  \bibitem{Maki_Par} A. B. Karki, Y. M. Xiong, N. Haldolaarachchige, S. Stadler, I. Vekhter, P. W. Adams, D. P. Young, W. A. Phelan, and J. Y. Chan, Phys. Rev. B 83, 144525 (2011).
  
  \bibitem{Coh_Leng} M. Tinkham, Introduction to Superconductivity, 2nd ed. (McGraw-Hill, New York, 1996)
  
   \bibitem{Pen_Dep} T. Klimczuk, F. Ronning, V. Sidorov, R. J. Cava, and J. D. Thompson, Phys. Rev. Lett. 99, 257004 (2007).
   
   \bibitem{McMillan} W. L. McMillan, Phys. Rev. 167, 331 (1968).
   
  \bibitem{R6H_WTRS} D. Singh, A. D. Hillier, A. Thamizhavel, and R. P. Singh, Phys. Rev. B 94, 054515 (2016).
  
  \bibitem{Y5R6S18} Amitava Bhattacharyya, Devashibhai Adroja, Naoki Kase, Adrian Hillier, Jun Akimitsu and Andre Strydom,Sci. Rep. 5, 12926 (2015).
 
 \bibitem{K2C3A3} D. T. Adroja, A. Bhattacharyya, M. Telling, Yu. Feng, M. Smidman, B. Pan, J. Zhao, A. D. Hillier, F. L. Pratt, and A. M. Strydom, Phys. Rev. B 92, 134505 (2015).
 
 \bibitem{SigSC_1} J. E. Sonier, J. H. Brewer, and R. F. Kiefl, Rev. Mod. Phys. 72, 769 (2000).

 \bibitem{SigSC_2} E. H. Brandt, Phys. Rev. B 37, 2349 (1988).
 
 \bibitem{L5R6S18} A. Bhattacharyya, D. T. Adroja, J. Quintanilla, A. D. Hillier, N. Kase, A. M. Strydom, and J. Akimitsu, Phys. Rev. B 91, 060503(R) (2015).
 
 \bibitem{ThCoC2} A. Bhattacharyya, D. T. Adroja, K. Panda, Surabhi Saha, Tanmoy Das, A. J. S. Machado, O. V. Cigarroa, T. W. Grant, Z. Fisk, A. D. Hillier, and P. Manfrinetti, Phys. Rev. Lett. 122, 147001 (2019).
 
 \bibitem{BP} Z. Sun, M. Enayat, A. Maldonado, C. Lithgow, E. Yelland, D. C. Peets, A. Yaresko, A. P. Schnyder, and P. Wahl, Nat. Commun. 6, 6633 (2015).

 \bibitem{KT} R. S. Hayano, Y. J. Uemura, J. Imazato, N. Nishida, T. Yamazaki, and R. Kubo, Phys. Rev. B 20, 850 (1979).

 \bibitem{CsCrAs} Devashibhai Adroja, Amitava Bhattacharyya, Michael Smidman, Adrian Hillier, Yu Feng, Bingying Pan, JunZhao, Martin R. Lees, Andre Strydom, and Pabitra K. Biswas, J. Phys. Soc. Jpn. 86, 044710 (2017).
 
 \bibitem{RRuB} J. A. T. Barker, R. P. Singh, A. D. Hillier, and D. McK. Paul, Phys. Rev. B 97, 094506 (2018).
 
 \bibitem{B22R} T Shang A Amon, D Kasinathan, W Xie, M Bobnar, Y Chen, A Wang, M Shi, M Medarde, H Q Yuan and T Shiroka, New J. Phys. 21 073034 (2019).
 
 \bibitem{Umera_Ref} D. A. Mayoh, J. A. T Barker, R. P. Singh, G. Balakrishnan, D. McK. Paul, and M. R. Lees, Phys. Rev. B 96, 064521 (2017).
 
 \bibitem{Unconv_1} K. Hashimoto, K. Cho, T. Shibauchi, S. Kasahara, Y. Mizukami, R. Katsumata, Y. Tsuruhara, T. Terashima, H. Ikeda, M. A. Tanatar, H. Kitano, N. Salovich, R. W. Giannetta, P. Walmsley, A. Carrington, R. Prozorov, and Y. Matsuda, Science 336, 1554 (2012).
 
 \bibitem{Unconv_2} R. Khasanov, H. Luetkens, A. Amato, H. H. Klauss, Z. A. Ren, J. Yang, W. Lu, and Z. X. Zhao, Phys. Rev. B 78, 092506 (2008).
 
 \bibitem{Umera} Y. J. Uemura, G. M. Luke, B. J. Sternlieb, J. H. Brewer, J. F. Carolan, W. N. Hardy, R. Kadono, J. R. Kempton, R. F. Kiefl, S. R. Kreitzman, P. Mulhern, T. M. Riseman, D. L. Williams, B. X. Yang, S. Uchida, H. Takagi, J. Gopalakrishnan, A. W. Sleight, M. A. Subramanian, C. L. Chien, M. Z. Cieplak, G. Xiao, V. Y. Lee, B. W. Statt, C. E. Stronach, W. J. Kossler, and X. H. Yu, Phys. Rev. Lett. 62, 2317 (1989).
 
 \bibitem{Tf} A. D. Hillier and R. Cywinski, Appl. Magn. Reson. 13, 95-109 (1997). 

\end{thebibliography}
\end{document}